\documentclass[aps,pre,reprint,superscriptaddress,longbibliography]{revtex4-1}
\usepackage{scrextend} %,wasysym}
\usepackage{amsmath,amssymb,graphicx,subfigure,color,times,tabularx,hyperref,fancyhdr}
\usepackage{esint}
\begin{document}
%\title{An exact on-grid potential and its associated particle mesh Ewald method}
\title{Infinite Boundary Terms and Pairwise Interactions: A Unified Framework for Periodic Coulomb Systems}
\author{Yihao Zhao}
\affiliation{Qingdao Institute for Theoretical and Computational Sciences (QiTCS), Center for Optics Research and Engineering, Shandong University, Qingdao 266237, P. R. China}
\author{Zhonghan Hu} \email{zhonghanhu@sdu.edu.cn}
\affiliation{Qingdao Institute for Theoretical and Computational Sciences (QiTCS), Center for Optics Research and Engineering, Shandong University, Qingdao 266237, P. R. China}
%\date{\today}
\begin{abstract} %The infinite boundary terms and the pairwise interactions introduced in the work entitled ``Infinite Boundary Terms of Ewald Sums and Pairwise Interactions for Electrostatics in Bulk and at Interfaces''[J. Chem. Theory Comput., 10, 5254, (2014)] 
The introduction of the infinite boundary terms and the pairwise interactions [J. Chem. Theory Comput., 10, 5254, (2014)] enables a physically intuitive approach for deriving electrostatic energy and pressure for both neutral and non-neutral systems under the periodic boundary condition (PBC).
For a periodic system consisting of $N$ point charges (with charge $q_j$ located at ${\mathbf r}_j$ where $j=1,2,\cdots N$) and one charge distribution of density $\rho({\mathbf r})$ within a primary cell of volume $V$, the derived electrostatic energy can be expressed as,
\[ {\mathcal U} = \sum_{i<j}^N q_iq_j\nu({\mathbf r}_{ij} ) + \sum_{j=1}^N q_j \int_V d{\mathbf r}_0\,\rho({\mathbf r}_0) \nu({\mathbf r}_{0j} ) + \frac{1}{2}\int_V d{\mathbf r}_0 \int_V d{\mathbf r}_1\,\rho({\mathbf r}_0)\rho({\mathbf r}_1) \nu({\mathbf r}_{01}),  \]
where ${\mathbf r}_{ij}={\mathbf r}_i - {\mathbf r}_j$ is the relative vector and $\nu({\mathbf r})$ represents the effective pairwise interaction under PBC.
The charge density $\rho({\mathbf r})$ is free of Delta-function-like divergence throughout the volume but may exhibit discontinuity.
This unified formulation directly follows that of the isolated system by replacing the Coulomb interaction $1/\lvert {\mathbf r} \rvert$ or other modified Coulomb interactions with $\nu({\mathbf r})$.
For a particular system of one-component plasma with a uniform neutralizing background, the implementation of various pairwise formulations clarifies the contribution of the background and subsequently reveals criteria for designing volume-dependent potentials that preserve the simple relation between energy and pressure.
\end{abstract} \maketitle
%%%%%%%%%%%%%%%%%%%%%%%%%%%%%%%%%%%%%%%%%%%%%%%%%%%%%%%%%%%%%%%%%%%%%%%%%%%%%%%%%%%%%%%%%%%%%%%%%%%%%%%%%%%%%%%%%%%%%%%%%%%%%%%%%%%%%%%%%%%%%%%%%%%%%%%%%%%%%%%%
\section{Introduction}
Since the seminal work of De Leeuw, Perram, and Smith\cite{DeLeeuw_Smith1980}, it has been recognized that the Coulomb lattice sum for electrically neutral systems of point charges is a conditionally convergent series whose value depends on the chosen order of summation.
The essence of this conditional convergence can be elucidated through a prototypical alternating series, \begin{equation} S = 1 - 1 + 1 - 1 + \cdots. \label{eq:s0} \end{equation}
Two summation conventions obviously exist:
\begin{equation} S_{-} = (1-1) + (1-1) + \cdots = 0, \end{equation}  leaving the last number at the boundary always $-1$, and \begin{equation} S_{+} = 1 + (-1 + 1) + (-1 + 1) + \cdots = 1, \label{eq:splus} \end{equation} leaving the last in the series always $+1$.
A unified representation, $S_{\mp} = S_{\rm bulk} \mp 1/2$, admits two contributions to the series: the bulk component satisfying the intrinsic periodicity, $S = 1 - S$, which gives $S=1/2 = S_{\rm bulk}$, and the boundary term, $\mp 1/2$, reflecting the influence of the chosen last number. 
The former can be alternatively expressed as an infinite geometric series: $S_{\rm bulk} = 1+x+x^2+\cdots = 1/(1-x)$, and then let $x=-1$.

The Coulomb lattice sum can be analyzed analogously. As the periodic lattice approaches infinity, distinct summation orders correspond to different geometries or symmetries of the lattice, each generating unique non-vanishing boundary term in general.
With the conditional boundary term removed, the remaining bulk component becomes well-defined and can be further expressed as a sum of two rapidly and absolutely convergent series via the Ewald technique, that was first developed by Paul Peter Ewald in 1921\cite{Ewald1921}.
The Ewald formulation, now commonly referred to as the three-dimensional Ewald summation associated with the tinfoil (conducting) boundary condition (e3dtf)\cite{DeLeeuw_Smith1980}, has become fundamental to both molecular dynamics simulations and electronic structure calculations of condensed-phase materials.

Although the Ewald summation method has been extensively revisited (e.g.\cite{Smith1988,Figueirido1995,Hummer1997,Hunenberger1999,Herce_Darden2007,Smith2008,Ballenegger2014,Remsing_Weeks2018}),
it remains highly nontrivial to generalize the method for deriving expressions of energy and pressure in complex periodic systems containing both point charges and charge distributions represented by a charge density.
The difficulty arises because, unlike the Coulomb energy of isolated systems, conventional Ewald formulations exhibit analytical complexity and lack a physically well-defined, easily generalizable pairwise decomposition.
Indeed, for the particular system of a one-component plasma with a uniform neutralizing background, inconsistency between the outputs of energy and pressure from the software LAMMPS\cite{lammps} was found\cite{Onegin2024}.
This inconsistency has been identified as the lack of a proper treatment of the background contribution\cite{Li2025,Demyanov2025}.
However, as remarked by Demyanov {\it et al.}\cite{Demyanov2025}, the proper incorporation of the background contribution remains unclear in maintaining thermodynamic consistency between energy and pressure calculations for systems governed by custom volume-dependent potentials under the periodic boundary condition (PBC).

The recently proposed framework of pairwise decomposition for periodic systems\cite{Hu2014ib} establishes a promising approach for overcoming the difficulty.
For a system of $N$ charged particles within a cubic cell of volume $V=L^3$, the particle-particle (pp) electrostatic energy under PBC/tinfoil boundary condition can be expressed as\cite{Hu2014ib,Yi_Hu2017pairwise,Pan_Hu2019,Hu2022}
\begin{equation} {\mathcal U}_{\rm pp} = \sum_{i<j}^N q_i q_j \nu_{\rm e3dtf}({\mathbf r}_{ij}, L),  \label{eq:Upp} \end{equation}
%\begin{equation} {\mathcal U}_{\rm pp} = \sum_{i<j}^N \nu_{\rm e3dtf}({\mathbf r}_{ij}, L) \equiv \frac{1}{2} \sum_{i=1}^N \sum_{\substack{j=1 \\ j\neq i}}^N \nu_{\rm e3dtf}({\mathbf r}_{ij}, L) ,  \label{eq:Upp} \end{equation}
where $q_j$ is the charge of the $j$-th particle and ${\mathbf r}_{ij}={\mathbf r}_i - {\mathbf r}_j$ is the displacement of the $i$-th particle from the $j$-th particle.
The pairwise e3dtf interaction, $\nu_{\rm e3dtf}({\mathbf r}, L)$, depends on the period $L$ and the vector ${\mathbf r}$, not merely on the radial distance $r=\lvert {\mathbf r} \rvert$. It exhibits even symmetry with respect to ${\mathbf r}$ and can be formally written as a simple Fourier series [e.g. Eqs. (5) of
Ref.\cite{Yi_Hu2017pairwise}]
\begin{equation} \nu_{\rm e3dtf}({\mathbf r}, L) = \frac{\xi}{L} + \lim_{P\to\infty} \sum_{ {\mathbf n}\neq {\mathbf 0}}^{P} \frac{e^{i 2\pi {\mathbf n}\cdot {\mathbf r}/L }}{\pi n^2 L} \quad\forall \, {\mathbf r}\in {\mathbb R}^3,  \label{eq:e3dtf}\end{equation}
with the constant $\xi$ being the ideal scattering coefficient for the simple cubic lattice: $\xi = 2.83\,729\,748\cdots$ (see Eq. (2.6) and Table I of Ref.\cite{Placzek1951}).
The physical meaning of $\xi$ will be explained alternatively in the present work.
In Eq.~\eqref{eq:e3dtf}, ${\mathbf n}$ stands for a triplet of integers: ${\mathbf n}=(n_1,n_2,n_3)\in {\mathbb Z}^3$ and $n=\lvert {\mathbf n}\rvert$.
For clarity, the double sum of an even function [Eq.~\eqref{eq:Upp}] and the triple sum of an arbitrary function [Eq.~\eqref{eq:e3dtf}] are both abbreviated throughout this paper,
\begin{equation}\sum_{i<j}^N \equiv \frac{1}{2} \sum_{i=1}^N \sum_{\substack{j=1\\j\neq i}}^N; \quad\quad \sum_{ {\mathbf n} \neq{\mathbf 0}}^P \equiv \sum_{ n_1 = - P }^{P}\, \sum_{ n_2 = - P}^{P}\, \sideset{}{'} \sum_{ n_3 = - P}^{P}, \label{eq:dtsum} \ \end{equation}
with the prime indicating that the ${\mathbf n}={\mathbf 0}$ term is excluded.

The pairwise formulation [Eq.~\eqref{eq:Upp}] shows that $\nu_{\rm e3dtf}({\mathbf r}, L) $ plays the same role in periodic systems as the Coulomb interaction $1/r$ does in isolated systems, regardless of whether the system is electrically neutral or not.
As such, it becomes straightforward to generalize the formulation to systems involving additionally a charge continuum described by a charge density, $\rho({\mathbf r})$,
\begin{equation} {\mathcal U}_{\rm pc} = \sum_{j=1}^N q_j \int_V d{\mathbf r}\, \rho({\mathbf r})\nu_{\rm e3dtf}({\mathbf r}-{\mathbf r}_j, L)  ,  \label{eq:Upc}  \end{equation}
for the particle-continuum (pc) interaction energy and
\begin{equation} {\mathcal U}_{\rm cc} = \frac{1}{2} \int_V d{\mathbf r} \int_V d{\mathbf r}^\prime \rho({\mathbf r})\rho({\mathbf r}^\prime) \nu_{\rm e3dtf}({\mathbf r}-{\mathbf r}^\prime, L)  , \label{eq:Ucc} \end{equation}
for the continuum-continuum (cc) interaction energy, respectively.
In scenarios where boundary terms alternative to the tinfoil boundary condition are imposed\cite{Hu2014ib}, the primary cell adopts non-cubic shape, or the basic interaction deviates from $1/r$,
the interaction $\nu_{\rm e3dtf}$ in Eqs.~\eqref{eq:Upp}-\eqref{eq:Ucc} is substituted with the corresponding effective pairwise interaction $\nu(\mathbf{r})$, and thus results in the generalized formula presented in the abstract.

The present unified pairwise formulation [Eqs.~\eqref{eq:Upp} to~\eqref{eq:Ucc}] for the general system consisting of both discrete point charges and a distribution of charge density does not seem to have appeared in literature.
Of course, an auxiliary Ewald splitting parameter can always be introduced to rewrite $\nu_{\rm e3dtf}({\mathbf r},L)$ exactly as a combination of two rapidly convergent series (see the Appendix or Eq.(3) of Ref.\cite{Yi_Hu2017pairwise}).
This combination differs from the pairwise Ewald potential $\nu_1({\mathbf r},L)+\nu_2({\mathbf r},L)$ proposed in Eqs.(2) and (3) of Ref.\cite{Demyanov2025} by Demyanov {\it et al.}. 
A key distinction is that $\nu_{\rm e3dtf}({\mathbf r},L)$, even in its computable form (e.g. Eq.(3) of Ref.\cite{Yi_Hu2017pairwise}), is independent of auxiliary parameters such as $\delta$ or $\alpha$\cite{Demyanov2025}, rendering it more physically meaningful.
%The corresponding Ewald formulation must be independent of the auxiliary parameter and therefore can be physically meaningful.
On the other hand, the concise expression [Eq.~\eqref{eq:e3dtf}], employed frequently in previous work\cite{Pan_Hu2019,Hu2022}, has proven useful for analytically predicting structural properties and dielectric response within the framework of the symmetry-preserving mean-field theory,
applicable to both interfacial\cite{Pan_Hu2017,Pan_Hu2019,Hu2014spmf} and bulk systems\cite{Pan_Hu2017,Hu2022}.
Stimulated by the recent fruitful discussions\cite{Li2025,Demyanov2025}, we apply various pairwise formulations to investigate the particular system of the one-component plasma with the uniform neutralizing background.
The main purposes of the present work are to emphasize the simplicity and usefulness of the unified framework, to illustrate more analytical properties associated with PBC, and to clarify the contribution of the background interactions.

The rest of this paper is organized as follows. Analogous to that done for Eqs.~\eqref{eq:s0} to~\eqref{eq:splus}, Section~\ref{sec:pair} conducts a step-by-step analysis of the lattice sum of two basic interactions, one being the usual $1/r$ and the other being a modified Coulomb interaction, also called the
angular-averaged (aa) Ewald potential\cite{Yakub_Ronchi2003,Yakub_Ronchi2005,Demyanov_Levashov2022,Demyanov_Levashov2022a}.
This analysis yields effective interactions, $\nu_{\rm e3dtf}({\mathbf r}, L)$ and $\nu_{\rm aa}({\mathbf r}, L)$, expressed as real- and Fourier-space series that rigorously incorporate the effect of PBC.
Building upon these series for $\nu_{\rm e3dtf}({\mathbf r}, L)$, $\nu_{\rm aa}({\mathbf r}, L)$, and other volume-dependent effective interactions, Section~\ref{sec:prop} elucidates universal properties collectively characterizing periodic electrostatic systems, including symmetry and positivity, lattice periodicity,
dominance over the Coulomb interaction, cancellation of electric field, constant average potential, constant potential of a uniform charge density, bulk invariance, and scaling behavior.
In Section~\ref{sec:ener}, we derive a unified pairwise formulation of electrostatic energies including Eqs.~\eqref{eq:Upp},~\eqref{eq:Upc} and~\eqref{eq:Ucc}. The effectiveness of these formulations is demonstrated by an example of calculating the Madelung constant for a crystal lattice.
For the one-component plasma with the uniform neutralizing background, our derivation produces results consistent with earlier work\cite{Li2025,Demyanov2025,Demyanov_Levashov2022,Demyanov_Levashov2022a} and we clarify that the electrostatic energy of the background must always be zero.
Section~\ref{sec:press} establishes thermodynamically consistent energy-pressure relations for systems governed by certain pairwise interactions, addressing the virial definition of pressure.
We discuss criteria for designing custom volume-dependent potentials such that the simple thermodynamic energy-pressure relation of the Coulomb system\cite{Li2025,Demyanov2025} can be maintained.
Concluding remarks are presented in Section~\ref{sec:con}. For completeness, conventional Ewald formulations and explicit expressions for other effective pairwise interactions are included in the Appendix.

%%%%%%%%%%%%%%%%%%%%%%%%%%%%%%%%%%%%%%%%%%%%%%%%%%%%%%%%%%%%%%%%%%%%%%%%%%%%%%%%%%%%%%%%%%%%%%%%%%%%%%%%%%%%%%%%%%%%%%%%%%%%%%%%%%%%%%%%%%%%%%%%%%%%%%%%%%%%%%%%

\section{Effective interactions under PBC}\label{sec:pair}
\begin{figure*}[!htb]\centerline{\includegraphics[width=16cm]{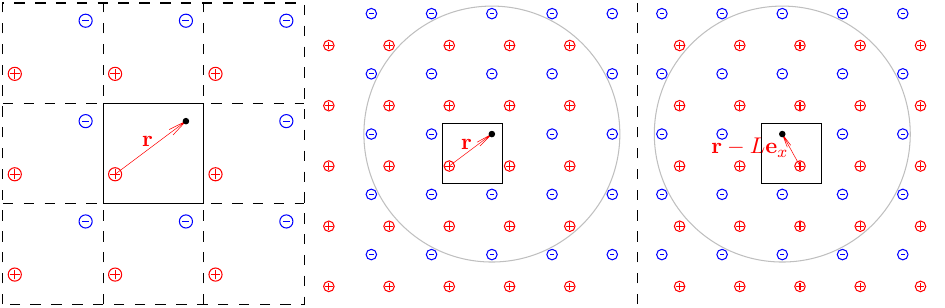} }           %%%%%%%%%%%%%%%%%%%%%%%%%%%%------------ Figure 1 ------------------------------------
\caption{A view of the cubic and even crystal in the $xy$-plane for $P=1$ (left) and $P=2$ (middle and right). The positive $x$-, $y$-, and $z$-axes point to the right, upward, and out of the screen or paper, respectively. The displacements, indicated by arrows, connect a source point charge to a target point.
For any $P$, the primary cell (solid square) containing the source charge and the target point is always located strictly at the center of the crystal.
When the target point is shifted by one period ($L$) in the negative $x$-direction, the surrounding charges remain unchanged in the bulk of the crystal (within the grey circle) but differ at the boundary. See Eq.~\eqref{eq:ubulk}.}\label{fig:crystal}\end{figure*}

The Coulomb interaction, $1/r$, describes the electric potential at a target point located at a displacement ${\mathbf r}$ from a unit point charge isolated in vacuum, that is, under the open boundary condition.
Under PBC, the unit source charge is placed inside a cubic cell with a length of $L$, and then this primary cell is replicated in all directions to form a perfect crystal.
The crystal contains not only the original source charge located inside the primary cell but also duplicated charges located inside all replicas.
If the target point at the displacement ${\mathbf r}$ is also located inside the primary cell, i.e., ${\mathbf r}\in[-L,L]^3$, the electric potential at the target point generated by the original source charge and all its duplicates must diverge as the number of replicas approaches infinity.
In order to eliminate this divergence, oppositely charged point charges are introduced at the displacement ${\mathbf r}$ from each duplicated charge, neutralizing all replicas.
Now, the electric potential generated by the crystal consisting of the source charge, all duplicates and their corresponding neutralizing charges, becomes finite and physically meaningful.

Perhaps surprisingly, given the displacement ${\mathbf r}$ and the period $L$, the electric potential of interest produced by an infinitely large crystal cannot be uniquely determined, although all replicas have been made electrically neutral.
In fact, the electric potential is composed of an intrinsic bulk term only depending on ${\mathbf r}$ and $L$, and an infinite boundary term additionally depending on the geometry of the macroscopic crystal relative to the primary cell.
The infinite boundary term for an arbitrary geometry is defined by\cite{Hu2014ib}
\begin{equation} \nu_{\rm ib}({\mathbf r}) = -\frac{2\pi}{V}\lim_{ {\mathbf k} \to {\mathbf 0} }  \frac{\left({\mathbf k}\cdot {\mathbf r} \right)^2}{k^2} . \label{eq:ib} \end{equation}
The orientation chosen for the conditional limit ${\mathbf k}\to {\mathbf 0}$ fully characterizes the summation order of the Coulomb lattice sum (see the Appendix and also Ref.\cite{Hu2014ib}). By definition,  $\nu_{\rm ib}({\mathbf r})$ is always negative for any ${\mathbf r}\neq {\mathbf 0}$.
The infinite boundary term expressed as the ${\mathbf k}\to {\mathbf 0}$ limit provides a simple way to help understand the vexing but important conditional convergence problem associated with PBC.
(e.g.\cite{Zhang2015,Antila2015,Lowe2018,Bakhshandeh_Levin2018,Urano_Yoshii2020,Ahrens-Iwers2021,Shi2022,Liang2023,Ai_Lomakin2024}).
More discussions of Eq.~\eqref{eq:ib} for crystals with fairly arbitrary geometries are given elsewhere.
At present, the replicas are assumed to distribute evenly around the primary cell such that the volume of the crystal grows according to $(2P+1)^3L^3$ with $P$ being an integer and $P\geqslant 1$. The three panels of Fig.~\ref{fig:crystal} show the cubic crystal in the $xy$ plane at $P=1$ and $P=2$.

To understand the bulk and boundary terms of the electric potential in such a highly symmetric geometry, we translate the target point by one period to create another displacement ${\mathbf r}-L{\mathbf e}_x \in[-L,L]^3$ and its corresponding cubic and even crystal in the right of Fig.1.
Obviously, when $P\to \infty$,  bulk charges surrounding the two target points can be overlapped, and then must produce an intrinsic bulk potential as a periodic function of the displacement.
On the other hand, surface charges at the infinite boundary differ and thereby produce the infinite boundary term, which depends on the relative geometry of the crystal, the period $L$, and the displacement.
Since the periodic translation from ${\mathbf r}$ to ${\mathbf r}-L{\mathbf e}_x $ yields different charge arrangements at the boundary, the infinite boundary term must not be a periodic function of the displacement. This non-periodic nature of the boundary term simply addresses the comments raised by Caillol\cite{Caillol1994}.

The cubic infinite boundary term corresponding to the present highly symmetric geometry reads explicitly\cite{Pan2017},
\begin{equation}  \nu_{\rm ib\_cub}({\mathbf r},L) = -\frac{2\pi}{L^3}\lim_{ {\mathbf k} ({\rm cub})\to {\mathbf 0} } \frac{\left({\mathbf k}\cdot {\mathbf r} \right)^2}{k^2} = -\frac{2\pi r^2}{3L^3},  \label{eq:cib} \end{equation}
which coincides with the spherical infinite boundary term evaluated previously\cite{Hu2014ib,Pan2017} and the spherical average of any infinite boundary term,
\begin{equation}  \frac{1}{4\pi} \int d\Omega\, \nu_{\rm ib}({\mathbf r}) = -\frac{2\pi}{V}\lim_{ {\mathbf k}\to {\mathbf 0} } \frac{ k^2 r^2 }{3k^2} = -\frac{2\pi r^2}{3V}.  \end{equation}
Here, $d\Omega = \sin\theta d\theta d\phi $ denotes the solid angle element, with $\theta$ and $\phi$ as the polar and azimuthal angles of the vector ${\mathbf r}$, respectively.

Given the boundary term expressed in the above explicit form, the remaining bulk component can be obtained by subtracting  $ \nu_{\rm ib\_cub}({\mathbf r},L) $ from the total electric potential generated by the crystal,
\begin{equation} \nu_P({\mathbf r}, L) = \frac{1}{r} + \sum_{ {\mathbf n} \neq{\mathbf 0}}^P \left( \frac{1}{\lvert {\mathbf r} -  {\mathbf n}L \rvert} - \frac{1}{nL} \right) + \frac{2\pi r^2}{3L^3}, \label{eq:ubulk} \end{equation}
where the periodic series accounts for the contribution from all replicas containing both the duplicated charges and the oppositely charged neutralizing charges.
When $P$ approaches infinity, $\nu_\infty({\mathbf r}, L)$ remains rigorously well-defined. It behaves like the Coulomb interaction at small distances but deviates at large distances.
In fact, the spherically averaged deviation is exclusively determined by the negative of the cubic infinite boundary term,
\begin{equation} \frac{1}{4\pi} \int d\Omega\, \nu_P({\mathbf r}, L) = \frac{1}{r} + \frac{2\pi r^2}{3L^3}, \quad\mbox{for}\quad  r \leqslant L, \label{eq:av} \end{equation}
since all terms in the periodic series vanish identically under spherical averaging,
\begin{equation} \frac{1}{4\pi} \int d\Omega\, \frac{1}{\left| {\mathbf r}-{\mathbf n}L \right|}  = \frac{1}{nL} \quad\mbox{for}\quad  nL \geqslant r .\end{equation}
This result is a direct manifestation of Newton's shell theorem\cite{Newton1687}---originally formulated for gravitational potentials---stating that, outside any spherically symmetric charge (mass) distribution, the potential is the same as if all the charge (mass) were concentrated at a point in the center.
%a spherically symmetric charge distribution produces the same external potential outside as a point charge at its center.

The lattice sum of $1/r$ defined in Eq.~\eqref{eq:ubulk} properly incorporates the effect of PBC while excluding the infinite boundary term. A similar definition applies to basic interactions other than $1/r$.
For example, one may truncate the spherical average of $\nu_P({\mathbf r}, L)$ at the radius of the sphere with the equivalent volume $4\pi r_{\rm s}^3/3 = V$.
The resultant modified Coulomb interaction, also called the angular-average (aa) Ewald potential, reads\cite{Yakub_Ronchi2003,Yakub_Ronchi2005,Demyanov_Levashov2022,Demyanov_Levashov2022a}
\begin{equation} w_{\rm aa}(r) = \frac{1}{r} + \frac{r^2}{2r_{\rm s}^3} - \frac{3}{2 r_{\rm s}} \quad \mbox{for}\quad r \leqslant r_{\rm s} =\frac{(3V)^{1/3}}{(4\pi)^{1/3}}, \label{eq:wr} \end{equation}
and $w_{\rm aa}(r) =0$ for $r > r_{\rm s}$. $w_{\rm aa}(r)$ approaches zero with zero derivative at $r_{\rm s}$. Its length scale is now characterized by $r_{\rm s}$.
Obviously, $w_{\rm aa}(\left|{\mathbf n}L \right|) = 0$ for any ${\mathbf n}\neq {\mathbf 0}$ and the lattice sum of $w_{\rm aa}(r)$ introduces no conditional convergence.
Therefore, the bulk component analogue to $\nu_\infty({\mathbf r}, L)$ in Eq.~\eqref{eq:ubulk} can be simply expressed as
\begin{equation} \nu_{\rm aa}({\mathbf r},L) = w_{\rm aa}(r) + \sum_{ {\mathbf n}\neq{\mathbf 0} }^\infty w_{\rm aa}(\lvert {\mathbf r} - {\mathbf n}L\rvert )  + \frac{3}{2 r_{\rm s}}, \label{eq:abulk} \end{equation}
where the constant $3/(2 r_{\rm s})$ has been added such that $\nu_{\rm aa}({\mathbf r},L)$ behaves like the Coulomb interaction at small distances as well.
%\textcolor{blue}{
As will be demonstrated in Section~\ref{sec:ener}, it is crucial to include this constant to guarantee the correct calculation of Madelung constants for crystals.
%}
Due to the short-range nature of $w_{\rm aa}(r)$, the summations over integers $n_1$, $n_2$, and $n_3$ in the above equation are constrained to at most three discrete values: typically $\{-1, 0, 1\}$ for ${\mathbf r}\in [-L,L]^3$.

The real-space expression [Eq.\eqref{eq:abulk}] can be converted to a Fourier-space expression via the well-known Poisson summation formula (e.g.\cite{Hu2014ib,Demyanov_Levashov2022})
\begin{equation} \nu_{\rm aa}({\mathbf r},L) = \frac{9}{5 r_{\rm s}} + \frac{1}{V}\sum_{ {\mathbf n}\neq{\mathbf 0} }^\infty  e^{i 2\pi {\mathbf n}\cdot {\mathbf r}/L }\hat{w}_{\rm aa}(2\pi{\mathbf n}/L) ,  \label{eq:wpbc}  \end{equation}
where $\hat{w}_{\rm aa}({\mathbf k})$ is the three-dimensional Fourier transform of $w_{\rm aa}(r)$,
\begin{equation} \hat{w}_{\rm aa}({\mathbf k}) = \frac{4\pi}{k^2}\left. \left[ 1 +  \frac{3k_{\rm s}\cos(k_{\rm s})-3\sin(k_{\rm s})}{k_{\rm s}^3} \right] \right|_{k_{\rm s} = kr_{\rm s}}. \end{equation}
At small $k$, $\hat{w}_{\rm aa}({\mathbf k})$ should determine long-range electrostatic correlations among charges according to the theory developed by one of us\cite{Hu2022,Gao_Hu2023}. 
Specifically, Eqs.(15) and (16) of Ref.\cite{Hu2022} predict the asymptotic behavior of the charge structure factor for conducting ionic fluids and insulating molecular fluids, respectively. 
However, their validity for systems interacting via the angular-averaged potential, where $\hat{\nu}({\mathbf k}|l,\alpha)=\hat{w}_{\rm aa}({\mathbf k})/(4\pi\varepsilon_0)$, remains to be confirmed by numerical simulations.
In the limit $k\to 0$ with $r_{\rm s}$ fixed, $\hat{w}_{\rm aa}({\mathbf k})$ approaches $2\pi r_{\rm s}^2/5$, leading to the constant term
\begin{equation} \frac{2\pi r_{\rm s}^2}{5}\frac{1}{L^3} + \frac{3}{2 r_{\rm s}}  = \frac{9}{5 r_{\rm s}}.  \end{equation}
On the other hand, in the limit $r_{\rm s}\to \infty$ followed by $k_{\rm s}\to \infty$,  $\hat{w}_{\rm aa}({\mathbf k})$ asymptotically converges to $4\pi/k^2$, which coincides with the Fourier transform of $1/r$. Consequently, Eq.~\eqref{eq:wpbc} in this limit differs from Eq.~\eqref{eq:e3dtf} solely by the constant term.
Direct validation  via the Poisson summation formula applied to $\nu_\infty({\mathbf r}, L)$ confirms
\begin{equation} \nu_\infty({\mathbf r}, L) = \nu_{\rm e3dtf}({\mathbf r},L) \quad\forall \, {\mathbf r}\in[-L,L]^3 , \label{eq:nu}\end{equation}
as rigorously derived in the Appendix.

Traditionally, the bare Coulomb interaction is often truncated at a fixed cutoff distance (denoted as $r_{\rm c}$), independent of both $V$ and $L$ (e.g.\cite{Wolf1999,Wu_Brooks2005,Fukuda2013,Wang_Fukuda2016,Hu2022}).
The resultant truncated interaction $w_{\rm cd}(r)$ might either decay sufficiently fast or smoothly taper to zero with a vanishing derivative at $r=r_{\rm c}$.
Using $w_{\rm cd}(r)$ as a basic interaction, the corresponding volume-dependent effective interaction under PBC can be expressed as a Fourier series
\begin{equation} \nu_{\rm cd}({\mathbf r},L) = \tau_{\rm cd} + \frac{1}{V}\sum_{ {\mathbf n}\neq{\mathbf 0} }^\infty  e^{i 2\pi {\mathbf n}\cdot {\mathbf r}/L }\hat{w}_{\rm cd}(2\pi{\mathbf n}/L) ,  \label{eq:wcdpbc}\end{equation}
where $\hat{w}_{\rm cd}({\mathbf k})$ is the Fourier transform of $w_{\rm cd}(r)$ and $\tau_{\rm cd}$ is a constant determined by enforcing that $\nu_{\rm cd}({\mathbf r},L) - 1/r$  strictly vanishes in the limit ${\mathbf r}\to{\mathbf 0}$.
Explicit forms of $w_{\rm cd}(r)$ and  $\nu_{\rm cd}({\mathbf r},L)$ for common truncation schemes are provided in the Appendix.
It has been analytically demonstrated that these short-ranged $w_{\rm cd}(r)$ developed in Refs.\cite{Wolf1999,Fukuda2013,Wang_Fukuda2016} fail to capture long-ranged charge-charge correlations\cite{Hu2022}. This limitation can be simply addressed through the symmetry-preserving mean-field approach\cite{Hu2022,Gao_Hu2023}.

%\textcolor{blue}{
The effective pairwise interaction defined in Eq.~\eqref{eq:wcdpbc} is periodic and does not vanish outside the primary cell. It is important to note that this effective pairwise interaction should not be confused with the truncated Coulomb interaction,
which vanishes beyond a cutoff distance but is often referred to as pair potentials or pairwise interactions in some literature (e.g.\cite{Wu_Brooks2005,Fukuda2013}).
%}
For both $w_{\rm aa}(r)$ and $w_{\rm cd}(r)$, their short-range nature ensures the unconditional ${\mathbf k}\to {\mathbf 0}$ limits of their Fourier transforms and the convergence of the Fourier series, Eqs.~\eqref{eq:wpbc} and~\eqref{eq:wcdpbc}, for any ${\mathbf r}$,
while the constants $\xi/L$, $9/(5r_{\rm s})$ and $\tau_{\rm cd}$ guarantee the correct short-range behavior of the effective interactions, $\nu_{\rm e3dtf}({\mathbf r},L)$, $\nu_{\rm aa}({\mathbf r},L)$, and $\nu_{\rm cd}({\mathbf r},L)$, respectively.

%\textcolor{blue}{
In the present work, we focus on systems with three-dimensional periodicity. However, extending the rigorous definition of the effective pairwise interaction to systems with two- (e.g.\cite{Smith2008,Hu2014ib,Pan_Hu2014,Gan2025}) or one-dimensional (e.g.\cite{Hu2014ib,Pan2023}) periodicity should be straightforward.
%}

%%%%%%%%%%%%%%%%%%%%%%%%%%%%%%%%%%%%%%%%%%%%%%%%%%%%%%%%%%%%%%%%%%%%%%%%%%%%%%%%%%%%%%%%%%%%%%%%%%%%%%%%%%%%%%%%%%%%%%%%%%%%%%%%%%%%%%%%%%%%%%%%%%%%%%%%%%%%%%%%
\section{Properties of the effective interactions}\label{sec:prop}  %h$\nu_{\rm e3dtf}$}\label{sec:prop}
\begin{figure}[!htb]\centerline{\includegraphics[width=7.7cm]{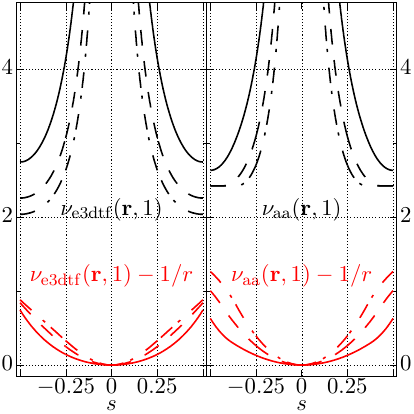} }          %---------------------  Figure -------------------------------------
\caption{The pairwise potentials $\nu({\mathbf r},1)$ (black in the top) generated by a unit charge at the origin and its difference from the Coulomb potential, $\nu({\mathbf r},1)-1/r$ (red at the bottom) along three typical directions: ${\mathbf r}=sL(1,0,0)$ (solid lines), ${\mathbf r}=sL(1,1,0)$ (dashed lines),
and ${\mathbf r}=sL(1,1,1)$ (dot-dashed lines).
Here, $s$ is a dimensionless variable and $L=1$ is the length of the cubic box centered at the origin. % and ${\mathbf r}$ represents the displacement from the center of the cubic cell with side length $L$.
Both potentials, $\nu_{\rm e3dtf}({\mathbf r},1)$ (left) and $\nu_{\rm aa}({\mathbf r},1)$ (right), exhibit flattening near the surface of the box. Similar behaviors can be found for $\nu_{\rm cd}({\mathbf r},L)$ in the Appendix.}\label{fig:e3dtf}\end{figure}

Before deriving the electrostatic energies, we examine the properties of the effective interactions, $\nu_{\rm e3dtf}({\mathbf r},L)$, $\nu_{\rm aa}({\mathbf r},L)$ and $\nu_{\rm cd}({\mathbf r},L)$, as formulated in both Fourier [Eqs.~\eqref{eq:e3dtf},~\eqref{eq:wpbc} and~\eqref{eq:wcdpbc}] and real [Eqs.~\eqref{eq:ubulk}
and~\eqref{eq:abulk}] spaces.
These complementary representations facilitate the concise derivation of exact results through algebraic operations such as substitution, differentiation, and integration.
The interactions $\nu_{\rm e3dtf}({\mathbf r},L)$, $\nu_{\rm aa}({\mathbf r},L)$ and $\nu_{\rm cd}({\mathbf r},L)$ share several fundamental properties that collectively characterize electrostatic systems with PBC.
For simplicity, we will henceforth use $\nu({\mathbf r},L)$ to generically denote any of these interactions---$\nu_{\rm e3dtf}({\mathbf r},L)$, $\nu_{\rm aa}({\mathbf r},L)$, or $\nu_{\rm cd}({\mathbf r},L)$---with specific distinctions provided where necessary.

(i) Symmetry and Positivity. $\nu({\mathbf r}, L)$ is even and strictly positive,
\begin{equation} \nu({\mathbf r},L) = \nu(-{\mathbf r},L) > 0 \quad \forall\, {\mathbf r}\in {\mathbb R}^3 .\end{equation}
The positivity of $\nu({\mathbf r}, L)$ imply that the effective interaction retains the long-range nature of the bare Coulomb interaction.
As is well known, this long-range characteristics leads to the divergence of the electrostatic energy for a homogeneous non-neutral system in the thermodynamic limit.

(ii) Lattice Periodicity. The interaction exhibits discrete translational symmetry,
\begin{equation} \nu({\mathbf r}, L) = \nu({\mathbf r} + {\mathbf n}L, L) \quad \forall\, {\mathbf n}\in {\mathbb Z}^3.  \end{equation}
Consequently, for any ${\mathbf r}\in {\mathbb R}^3$, there exists a minimum image ${\mathbf r}_m \in [-L/2,L/2]^3$ such that,
\begin{equation} \nu({\mathbf r}_m,L) =  \nu({\mathbf r},L),  \end{equation}
which allows mapping any interaction to the case where the source charge is located at the center of the primary cell and the target point lies within the cell.

(iii) Dominance over the Coulomb interaction.
\begin{equation} \nu({\mathbf r}, L) \geqslant 1/r \quad\forall\, {\mathbf r}\in {\mathbb R}^3 . \end{equation}
By property (ii), it suffices to verify this relation for ${\mathbf r}\in [-L/2,L/2]^3$.
Fig~\ref{fig:e3dtf} reveals that the difference, $\nu({\mathbf r},L) - 1/r$, along typical directions approaches $0$ at ${\mathbf r}\to {\mathbf 0}$ and reaches its maximum at the surface of the primary cell.

(iv) Cancellation of Electric Field. The electric field at ${\mathbf r}=(x,y,z)$ in the $x$, $y$, or $z$ direction vanishes at the surface normal to the direction,
\begin{equation}  - {\mathbf e}_x \cdot \nabla \nu({\mathbf r}, L) = 0 \quad\mbox{for}\quad x = \pm L/2,  \label{eq:ve} \end{equation}
with analogous relations for $y$ and $z$. This cancellation of the electric field can be rigorously proved using the Fourier-space expressions given in Eqs.~\eqref{eq:e3dtf}, ~\eqref{eq:wpbc} and~\eqref{eq:wcdpbc} (see the Appendix).
It implies the symmetric arrangement of charges around the target point in one or more particular directions.

(v) Constant Average Potential.
\begin{equation} \frac{1}{V} \int_V d{\mathbf r}\, \nu({\mathbf r} - {\mathbf r}_0,L) = \tau([\nu]) ,\label{eq:cap} \end{equation}
where $\tau([\nu])$ is independent of ${\mathbf r}_0$ and is the constant term of the Fourier-space expressions, i.e.,
\begin{equation} \tau([\nu_{\rm e3dtf}])=\frac{\xi}{L};\quad\tau([\nu_{\rm aa}])=\frac{9}{5r_{\rm s}};\quad \tau([\nu_{\rm cd}])=\tau_{\rm cd}. \label{eq:tau} \end{equation}
This constancy originates from the orthogonality of Fourier series. Specifically, the integral of any non-constant trigonometric function (e.g. $\sin(2\pi n_1 x/L)$ or $\cos(2\pi n_1x/L)$) over any length of the period $L$ vanishes identically,
%\begin{equation}
\begin{multline} \int_{x_0}^{x_0+L}dx\, \sin\frac{2\pi n_1 x}{L} = \int_{x_0}^{x_0+L}dx\, \cos\frac{2\pi n_1 x}{L} = 0  \\ \quad\forall n_1\in {\mathbb Z}\quad\mbox{and}\quad  n_1\neq 0  \label{eq:vtri} ,\end{multline}
%\end{equation}
leaving only the constant component in the Fourier expansion of $\nu({\mathbf r},L)$ [Eqs.~\eqref{eq:e3dtf},~\eqref{eq:wpbc} and~\eqref{eq:wcdpbc}].
In the context of the symmetry-preserving mean-field theory, Eq.~\eqref{eq:vtri} enables systematic reduction of degrees of freedom while preserving the intrinsic symmetries of the system\cite{Pan_Hu2019,Hu2022}.

(vi) Constant Potential of a Uniform Charge Density. A rewrite of Eq.~\eqref{eq:vtri} gives
\begin{equation} \int_V d{\mathbf r}\, \frac{1}{V} \nu({\mathbf r}_0 - {\mathbf r},L) = \tau([\nu]) .\label{eq:cpu} \end{equation}
Eqs.~\eqref{eq:cap} and~\eqref{eq:cpu} offer clear physical interpretations of the constant $\tau([\nu])$.
$\xi/L$ represents either the bulk potential averaged over the primary cell, generated by a unit point charge located at an arbitrary position, or the bulk potential at an arbitrary point, generated by a uniform charge density $1/V$, under PBC.
The other two constants, $9/(5r_{\rm s})$ and $\tau_{\rm cd}$, follow analogously.

Although the three effective interactions share properties (i)-(vi), their underlying basic interactions---$1/r$, $w_{\rm aa}(r)$ and $w_{\rm cd}(r)$---differ fundamentally.
Notably, $w_{\rm aa}(r)$ by itself depends on the system's periodicity through the parameter $r_{\rm s}$, whereas the bare Coulomb interaction and $w_{\rm cd}(r)$ do not.
Consequently, the bulk potential described by $\nu_{\rm aa}({\mathbf r},L)$ depends not only on the configuration of the surrounding charges but also on the setup of the primary cell.
In contrast, the following property of bulk invariance applies exclusively to $\nu_{\rm e3dtf}({\mathbf r},L)$ and $\nu_{\rm cd}({\mathbf r},L)$.

(vii) Bulk Invariance.
In the left panel of Fig.~\ref{fig:crystal} ($P=1$),  if all the charges within the crystal collectively act as the source subject to an extended periodicity of $3L$,  the potential at the target point remains unchanged, provided the basic interaction is $L$-independent,
\begin{equation} \nu({\mathbf r},L) = \nu({\mathbf r},3L) + \sum_{ {\mathbf n}\neq{\mathbf  0} }^1 \left[ \nu( {\mathbf r}+{\mathbf n}L, 3L) - \nu({\mathbf n}L, 3L ) \right]. \label{eq:inva}  \end{equation}
This invariance holds even when generalized to arbitrary primary cells with variations in shape, size, or both.

Conversely, defining $r=sL$ reveals that both $1/(sL)$ and $w_{\rm aa}(sL)$ are inversely proportional to $L$ for the fixed dimensionless parameter $s$,  whereas $w_{\rm cd}(sL)$ is not.
As a consequence, the following property of scaling behavior applies exclusively to $\nu_{\rm e3dtf}({\mathbf r},L)$ and $\nu_{\rm aa}({\mathbf r}, L)$.

(viii) Scaling Behavior. For any fixed dimensionless vector ${\mathbf s}$, $\nu({\mathbf s}L, L)$ scales as $1/L$, yielding an identity of the derivative:
\begin{equation} \left. \frac{\partial \nu({\mathbf s}L, L) }{\partial L}  \right|_{ {\mathbf s} } = - \frac{\nu({\mathbf s}L, L) }{L} .  \label{eq:sb} \end{equation}
This scaling behavior underpins the classical result that the pressure of a Coulomb system is uniquely determined by its potential energy---a relation known since the 19th century \cite{Onegin2024}.
In the field of molecular dynamics simulations, this energy-pressure relation often serves as a validation criterion for the convergence of the Ewald summation (e.g.\cite{Smith2002,dlpoly}).
Eq.~\eqref{eq:sb} establishes critical criteria for effective interactions to preserve the energy-pressure relation, as detailed in Section~\ref{sec:press}.

%%%%%%%%%%%%%%%%%%%%%%%%%%%%%%%%%%%%%%%%%%%%%%%%%%%%%%%%%%%%%%%%%%%%%%%%%%%%%%%%%%%%%%%%%%%%%%%%%%%%%%%%%%%%%%%%%%%%%%%%%%%%%%%%%%%%%%%%%%%%%%%%%%%%%%%%%%%%%%%%
\section{Electrostatic Potentials and Energies}\label{sec:ener}
\begin{figure}[!htb]\centerline{\includegraphics[width=7cm]{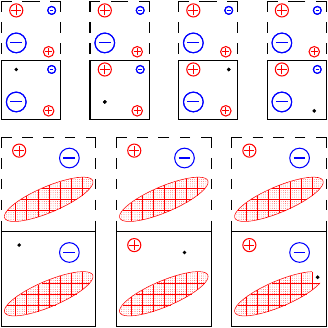} }     %%%%%%%%%%%%%%%%%%%%%%%%%%%%------------ Figure ------------------------------------
\caption{Primary cells (solid squares) and their replicas (dashed squares) for a system composed of pure point charges (top) and a system composed of both point charges and a charge distribution (bottom).
The target point is located either at the position of a point charge or at an infinitesimally small region (bottom right) of the charge continuum.
The corresponding source is always the collection of the remaining charges. As long as the system is electrically neutral, the replica remains identical to the system, regardless of the target point. See Eqs.~\eqref{eq:phipp}, ~\eqref{eq:phip} and ~\eqref{eq:phic}. }\label{fig:cont}\end{figure}

The effective interaction $\nu({\mathbf r}, L)$ corresponds to $\nu_{\rm e3dtf}({\mathbf r}, L)$ for the Coulomb interaction,  $\nu_{\rm aa}({\mathbf r}, L)$ for the angular-averaged interaction truncated at the $L$-dependent distance $r_{\rm s}$,
and $\nu_{\rm cd}({\mathbf r}, L)$ for some modified Coulomb interaction truncated at the fixed cutoff $r_{\rm c}$, respectively.
It represents the bulk electric potential produced by a unit charge under PBC, expressed in a unified manner.
Under this unified framework, a point charge $q$, acting as the source, generates an electric potential $q\nu_{\rm }({\mathbf r},L)$ at the displacement ${\mathbf r}$, analogous to the conventional electric potential $q/r$ generated under the open boundary condition.
This analysis extends naturally to any $N$-particle system. When focusing on the position of the $i$-th particle ${\mathbf r}_i$, the remaining $N-1$ charges collectively act as the source, generating an electric potential at ${\mathbf r}_i$:
\begin{equation} \phi_{\rm pp}(i) = \sum_{ j=1,j\neq i }^N q_j \nu_{\rm }({\mathbf r}_i - {\mathbf r}_j,L).  \label{eq:phipp} \end{equation}
For charge-neutral systems satisfying the constraint, $\sum_{j=1}^N q_j=0$, the neutralizing charge in any replica must equal $q_i$, since it is negative to the sum of the $N-1$ charges:
\begin{equation} -\sum_{j=1,j\neq i}^N q_j = q_i.  \end{equation}
As shown in the top panel of Fig.~\ref{fig:cont}, this constraint ensures that all replicas remain identical to the original $N$-particle system regardless of which particle is focused.
Consequently, $\phi_{\rm pp}(i)$ (for any $i=1,2,\cdots,N$) fully accounts for interactions with replicas of the entire $N$-particle system.

If the system includes an extra charge continuum treated as a collection of infinitely many point charges, each contributing an infinitesimal charge element $\rho({\mathbf r}) d{\mathbf r}$ at ${\mathbf r}$, the total electric potential experienced by the $i$-th point charge becomes
\begin{equation} \phi_{\rm p}(i)  = \phi_{\rm pp}(i) + \phi_{\rm pc}({\mathbf r}_i) , \label{eq:phip} \end{equation}
where $\phi_{\rm pc}({\mathbf r})$ is the electric potential at ${\mathbf r}$ produced solely by the charge distribution,
\begin{equation} \phi_{\rm pc}({\mathbf r}) = \int_V d{\mathbf r}^\prime\, \rho({\mathbf r}^\prime) \nu_{\rm }({\mathbf r} - {\mathbf r}^\prime,L). \label{eq:phipc}  \end{equation}
Similarly, the total electric potential experienced by the charge element at ${\mathbf r}$ is given by
\begin{equation} \phi_{\rm c}({\mathbf r})  = \sum_{j=1}^N q_j \nu_{\rm }({\mathbf r} - {\mathbf r}_j,L) + \phi_{\rm pc}({\mathbf r}).  \label{eq:phic} \end{equation}
Again, when the system is electrically neutral, satisfying
\begin{equation} \int_V d{\mathbf r}\, \rho({\mathbf r}) + \sum_{j=1}^N q_j = 0, \end{equation}
the potentials $\phi_{\rm p}(i)$ and $\phi_{\rm c}({\mathbf r})$ both account for interactions with replicas of the entire neutral system, as illustrated in the bottom panel of Fig.~\ref{fig:cont}.

In the formulations above, point charges and the distribution of a charge density are treated separately.
While discrete charges can, in principle, be represented as Dirac delta functions and combined with $\rho({\mathbf r)}$ to form the total charge density, the resulting electric potential generated by this combined density diverges precisely at the location of each discrete charge.
Eliminating this divergence would complicate the mathematical treatment.
Instead, Eqs.~\eqref{eq:phipp},~\eqref{eq:phip}, and~\eqref{eq:phic} naturally avoid these divergences, making them more convenient for deriving electrostatic energies in systems containing both point charges and charge distributions.

The electric potentials acting on the point charges and the charge continuum yield the electrostatic energy of all point charges,
\begin{equation} {\cal U}_{\rm p} = \frac{1}{2}\sum_{i=1}^N q_i  \phi_{\rm p}(i) , \label{eq:Up} \end{equation}
and that of the charge continuum,
\begin{equation} {\cal U}_{\rm c} = \frac{1}{2} \int_V d{\mathbf r}\, \rho({\mathbf r}) \phi_{\rm c}({\mathbf r}) , \label{eq:Uc} \end{equation}
respectively.
After substituting the expressions for $\phi_{\rm p}(i)$ and $\phi_{\rm c}({\mathbf r})$ from Eq.~\eqref{eq:phipp} and Eqs.~\eqref{eq:phip} to~\eqref{eq:phic}, the total electrostatic energy can alternatively be expressed as
\begin{equation} {\cal U} =  {\cal U}_{\rm p} +  {\cal U}_{\rm c} = {\cal U}_{\rm pp} +  {\cal U}_{\rm pc} +  {\cal U}_{\rm cc},  \label{eq:totalU} \end{equation}
where the particle-particle (pp), particle-continuum (pc), and continuum-continuum (cc) interaction energies are given by
\begin{equation} {\cal U}_{\rm pp} = \sum_{i<j}^N q_i q_j \nu({\mathbf r}_{ij}, L)   , \label{eq:Upp2} \end{equation}
\begin{equation}  {\mathcal U}_{\rm pc} = \sum_{j=1}^N q_j \int_V d{\mathbf r}\, \rho({\mathbf r})\nu({\mathbf r}-{\mathbf r}_j, L)  ,  \label{eq:Upc2} \end{equation}
and
\begin{equation}  {\mathcal U}_{\rm cc} = \frac{1}{2} \int_V d{\mathbf r} \int_V d{\mathbf r}^\prime \rho({\mathbf r})\rho({\mathbf r}^\prime) \nu({\mathbf r}-{\mathbf r}^\prime, L)  , \label{eq:Ucc2} \end{equation}
respectively.
Eqs.~\eqref{eq:totalU} to~\eqref{eq:Ucc2} identify with the equation provided in the abstract and reduce to Eqs.~\eqref{eq:Upp}, ~\eqref{eq:Upc} and~\eqref{eq:Ucc} upon setting $\nu = \nu_{\rm e3dtf}$.

Eq.~\eqref{eq:Up} to~\eqref{eq:Ucc2} provides a unified framework for energies of periodic Coulomb systems, where the effective pairwise interaction is described by $\nu({\mathbf r}, L)$.
For the one-component plasma of $N$ identical charges, $q_1=\cdots=q_N=q_0$, with the uniform neutralizing background, the energies simplify as follows
\begin{equation} {\mathcal U}_{\rm c}=0; \quad {\mathcal U}_{\rm pc} = - 2\, {\mathcal U}_{\rm cc} = -\tau([\nu]) N^2 q_0^2,  \label{eq:ocp} \end{equation}
which arises from the electro-neutrality condition and properties (v) and (vi) discussed in the preceding section.
The combination of Eqs.~\eqref{eq:totalU},~\eqref{eq:Upp2}, and~\eqref{eq:ocp}, along with the specific choices of  $\nu=\nu_{\rm e3dtf}$ (see the Appendix), $\nu=\nu_{\rm aa}$  [Eq.~\eqref{eq:abulk}], and $\tau([\nu])$ [Eq.~\eqref{eq:tau}], is fully consistent with previous results reported by Li {\it et al.}\cite{Li2025},
Onegin {\it et al.}\cite{Onegin2024}, and Demyanov {\it et al.}\cite{Demyanov2025}.
To be more precise, Eq.~\eqref{eq:totalU} identifies with Eq.(29) of Ref.\cite{Onegin2024} for $\nu = \nu_{\rm aa}$ and both Eq.(3.18) of Ref.\cite{Li2025} and Eq.(1) of Ref.\cite{Demyanov2025} for $\nu = \nu_{\rm e3dtf}$.
Note that Eq.(3.19) and Eq.(3.20) of Ref.\cite{Li2025} contains a typo: $\alpha$ should be replaced by $\sqrt{\alpha}$. The inconsistency between $\alpha$ and $\sqrt{\alpha}$ was already pointed out by Demyanov {\it et al.} in Ref.\cite{Demyanov2025}.
Hence, the consistency between Eq.~\eqref{eq:totalU} and the existing expressions for the one-component plasma well demonstrates the simplicity and versatility of the present unified framework.

\begin{figure}[!htb]\centerline{\includegraphics[width=7cm]{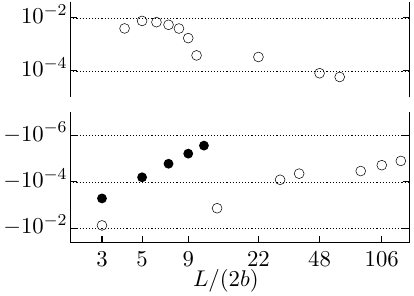} }          %---------------------  Figure 4-------------------------------------
\caption{
%\textcolor{blue}{
Logarithmic plot of the deviations of the Madelung constants from the exact value ($1.74\,756\,459\cdots$). These Madelung constants as a function of $L$ are computed via Eqs.~\eqref{eq:m},~\eqref{eq:abulk}, and~\eqref{eq:direct} and listed in Tabs.~\ref{tab:aa} and~\ref{tab:direct}.
Open and filled circles correspond to the data in Tabs.~\ref{tab:aa} and~\ref{tab:direct}, respectively.
%}
}\label{fig:error} \end{figure}

\begin{table}[!htb] \begin{center}\caption{Coordinates of ions in a cubic unit cell with length $2b$, where $b$ is the bond length ($b=2.789$\AA\, for NaCl\cite{Ghate1965,Shi_Wortis1988}). The face-centered cubic (fcc) crystal is terminated by six $(100)$ crystallographic planes.
This unit cell has zero net charge, zero dipole moment, and zero quadrupole moment.
Note: the value of $b$ is provided for contextual reference but is not directly used in calculations via Eqs.~\eqref{eq:mexact},~\eqref{eq:m} and~\eqref{eq:direct}.} \label{tab:unit}
\begin{tabular}{ccc} \hline  cell   &  \quad$4$ Na$^+$            &      \quad$4$ Cl$^-$         \\[0.5ex]
 $ \begin{aligned} L=2b \\ N = 8\end{aligned} $    & \quad  $\begin{aligned} (0,0,0&)(b,0,b) \\ (b,b,0&)(0,b,b) \end{aligned}$ \quad  &  $\begin{aligned} (b,0,0&) (0,0,b) \\  (b,b,b&) (0,b,0) \end{aligned}$  \\ \hline
\end{tabular} \end{center} \end{table}
\begin{table}[!htb]
\caption{Madelung constant computed via Eqs.~\eqref{eq:m} and~\eqref{eq:abulk} for cubic primary cells of length $L$, containing $N=L^3/b^3$ ions. $N_{\rm s}$ denotes the number of ions within a sphere of radius $r_{\rm s}=(3/(4\pi))^{1/3} L$ centered at any reference ion.
The exact Madelung constant is $1.74\,756\,459\cdots $. See also Table 2 of Ref.~\cite{Demyanov_Levashov2022} and Table I of Ref.~\cite{Yakub_Ronchi2003}.} \label{tab:aa}
\begin{tabular}{lrrcc} \hline  $L/(2b)$   &  $N_{\rm s} - N $\quad\quad       &  Net Charges                         &   $M$   &  Difference (\%) \\[0.3ex] \hline
                               $ 1   $    &     \mbox{  -1\quad\quad}     &    \mbox{   -5\quad\quad}            &    1.525826        & \mbox{-12.6884}  \\[0.3ex] \hline
                               $ 2   $    &     \mbox{  17\quad\quad}     &    \mbox{    5\quad\quad}            &    1.716726        &        -1.76465  \\[0.3ex] \hline
                               $ 3   $    &     \mbox{ -13\quad\quad}     &    \mbox{  -29\quad\quad}            &    1.739927        &        -0.43706  \\[0.3ex] \hline
                               $ 4   $    &     \mbox{ -27\quad\quad}     &    \mbox{   13\quad\quad}            &    1.751516        & \mbox{ 0.22613}  \\[0.3ex] \hline
                               $ 5   $    &     \mbox{  21\quad\quad}     &    \mbox{   41\quad\quad}            &    1.755085        & \mbox{ 0.43035}  \\[0.3ex] \hline
                               $ 6   $    &     \mbox{  15\quad\quad}     &    \mbox{   31\quad\quad}            &    1.754329        & \mbox{ 0.38707}  \\[0.3ex] \hline
                               $ 7   $    &     \mbox{  33\quad\quad}     &    \mbox{   41\quad\quad}            &    1.752962        & \mbox{ 0.30885}  \\[0.3ex] \hline
                               $ 8   $    &     \mbox{ -29\quad\quad}     &    \mbox{  119\quad\quad}            &    1.751490        & \mbox{ 0.22461}  \\[0.3ex] \hline
                               $ 9   $    &     \mbox{ -89\quad\quad}     &    \mbox{   55\quad\quad}            &    1.749271        & \mbox{ 0.09766}  \\[0.3ex] \hline
                              %$ 9\footnote{The direct sum gives the Madelung constant 1.747558 with a relative difference of -0.00035\%, which is more accurate than the last.}$    &     \mbox{ -89\quad\quad}     &    \mbox{   55\quad\quad}            &    1.749271        & \mbox{ 0.09766}  \\[0.3ex] \hline
                               $ 10  $    &     \mbox{  25\quad\quad}     &    \mbox{  -31\quad\quad}            &    1.747946        & \mbox{ 0.02185}  \\[0.3ex] \hline
                               $ 13  $    &     \mbox{ -19\quad\quad}     &    \mbox{    5\quad\quad}            &    1.746176        &       -0.07948   \\[0.3ex] \hline
                               $ 22  $    &     \mbox{ -25\quad\quad}     &    \mbox{   -1\quad\quad}            &    1.747898        & \mbox{ 0.01906}  \\[0.3ex] \hline
                               $ 29  $    &     \mbox{  55\quad\quad}     &    \mbox{   55\quad\quad}            &    1.747483        &       -0.00469   \\[0.3ex] \hline
                               $ 37  $    &     \mbox{ 255\quad\quad}     &    \mbox{  -77\quad\quad}            &    1.747520        &       -0.00254   \\[0.3ex] \hline
                               $ 48  $    &     \mbox{ 435\quad\quad}     &    \mbox{  379\quad\quad}            &    1.747647        & \mbox{ 0.00470}  \\[0.3ex] \hline
                               $ 62  $    &     \mbox{-239\footnote{\label{error1}These two numbers, -239 and 1699, differ from the corresponding numbers, -230 and 1700, in Table 2 of Ref.\cite{Demyanov_Levashov2022}.}\quad\quad}  &    \mbox{   25\quad\quad}    &    1.747624        & \mbox{ 0.00339}  \\[0.3ex] \hline
                               $ 81  $    &     \mbox{ 333\quad\quad}     &    \mbox{-1427\quad\quad}        &    1.747530        &       -0.00196   \\[0.3ex] \hline
                               $ 106 $    &     \mbox{-277\quad\quad}     &    \mbox{ -829\quad\quad}        &    1.747545        &       -0.00110   \\[0.3ex] \hline
                               $ 135 $    &     \mbox{1699\footref{error1}\quad\quad}     &  \mbox{ -293\quad\quad}    &    1.747552        &       -0.00072  \\[0.0ex] \hline
\end{tabular}
\end{table}
To illustrate the difference between $\nu_{\rm e3dtf}$ and $\nu_{\rm aa}$, we compute the Madelung constant ($M$), which is defined as the electric potential experienced by any given charge in a crystal, $\phi_{\rm crystal}(i)$, relative to that generated solely by its neighbors in a gas phase, $\phi_{\rm gas}(i)$.
For the NaCl crystal, whose unit cell is described in Tab.~\ref{tab:unit}, $\phi_{\rm gas}(i)$ arises from the nearest counter ion located at a distance of $b$, yielding $\phi_{\rm gas}(i) = -q_i /b$. The exact Madelung constant is given by
\begin{equation}  M = -\frac{b}{q_i} \sum_{\substack{j=1\\j \neq i}}^8  q_j \nu_\infty({\mathbf r}_i - {\mathbf r}_j, 2b ) =1.74\,756\,459\cdots, \label{eq:mexact} \end{equation}
where $\nu_\infty$ is defined in Eq.~\eqref{eq:ubulk} with $P=\infty$. The Madelung constant $M$ is independent of $i$ because all ions occupied equivalent positions in the crystal lattice.
Notably, the $r^2$ term of Eq.~\eqref{eq:ubulk} does not contribute to $M$, as shown by the following identity,
\begin{equation} \sum_{j=1}^8 q_j \left|{\mathbf r} - {\mathbf r}_j \right|^2 = \sum_{j=1}^8 q_j \left( {\mathbf r}_j\cdot {\mathbf r}_j - 2 {\mathbf r}\cdot{\mathbf r}_j + r^2\right) = 0, \end{equation}
which holds for any ${\mathbf r}$ because the total charge and the dipole and quadrupole moments of the unit cell all vanish.
Furthermore, by virtue of the bulk invariance [property (vii) and Eq.~\eqref{eq:inva}], the Madelung constant remains invariant regardless of whether a larger or smaller primary cell is used.
In contrast, if the underlying basic interaction is replaced by the angular-averaged interaction $w_{\rm aa}(r)$, the computed Madelung constant,
\begin{equation}  M = -\frac{b}{q_i} \sum_{\substack{j=1\\j \neq i}}^N  q_j \nu_{\rm aa}({\mathbf r}_j - {\mathbf r}_i, L), \label{eq:m} \end{equation}
exhibits a strong dependence on the size of the primary cell, as demonstrated in Tab.~\ref{tab:aa} and Fig.~\ref{fig:error}.
Eq.~\eqref{eq:m} appears to converge to the exact value for extremely large primary cells but the convergence is slow. Alternatively, one might directly compute the Madelung constant using the geometry of Fig.~\ref{fig:crystal}, as follows,
\begin{equation}  M = -\frac{b}{q_i} \sum_{\substack{j=1\\j \neq i}}^N  \frac{q_j}{\left| {\mathbf r}_j - {\mathbf r}_i \right| } ,\label{eq:direct} \end{equation}
where the boundary term is omitted since it makes no contribution for the present unit cell. As shown in Tab.~\ref{tab:direct} and Fig.~\ref{fig:error}, Eq.~\eqref{eq:direct} converges significantly faster than Eq.~\eqref{eq:m}.
For instance, with $P=4$ corresponding to a crystal size of $L/(2b) = 9$, the Madelung constant computed via Eq.~\eqref{eq:direct} is already more accurate than that obtained via Eq.~\eqref{eq:m} at $L/(2b) = 135$.
\begin{table}[!htb]
\caption{Madelung constant computed via Eq.~\eqref{eq:direct} for cubic primary cells of length $L$. The net charges are always zero.}\label{tab:direct}
\begin{tabular}{lcccc}  \hline $L/(2b)$  & \quad$P$\quad\quad   & $N$           &  $M$                &  Difference (\%) \\[0.3ex] \hline
                               1         &  0                   & 8             &        1.456030     &  -16.6823  \\ [0.3ex] \hline
                               3         &  1                   & 216           &        1.747042     &   -0.02993  \\ [0.3ex] \hline
                               5         &  2                   & 1000          &        1.747501     &   -0.00367  \\ [0.3ex] \hline
                               7         &  3                   & 2744          &        1.747548     &   -0.00096 \\ [0.3ex] \hline
                               9         &  4                   & 5832          &        1.747558     &   -0.00035 \\ [0.3ex] \hline
                               11        &  5                   & 10648         &        1.747562     &   -0.00016  \\ [0.3ex] \hline

\end{tabular}
\end{table}

%%%%%%%%%%%%%%%%%%%%%%%%%%%%%%%%%%%%%%%%%%%%%%%%%%%%%%%%%%%%%%%%%%%%%%%%%%%%%%%%%%%%%%%%%%%%%%%%%%%%%%%%%%%%%%%%%%%%%%%%%%%%%%%%%%%%%%%%%%%%%%%%%%%%%%%%%%%%%%%%
\section{Pressure}\label{sec:press}
In the Boltzmann-Gibbs framework of statistical mechanics, the canonical partition function $Q$ of the periodic system in its quasi-classical form can be factorized into two components: an ideal gas component ($Q_{\rm id}$) and an excess configurational integral ($Q_{\rm ex}$),
\begin{equation} \begin{aligned} Q(N,V,T) &= Q_{\rm id}(N,V,T) \cdot Q_{\rm ex}(N,V,T) \\ & = \frac{V^N}{\Lambda^{3N} N! } \cdot \frac{1}{V^N} \int_0^L d\bar{\mathbf r}\, e^{-\beta {\mathcal U}(\bar{\mathbf r}, L)} \end{aligned}\quad , \label{eq:can} \end{equation}
where the thermal de Broglie wavelength is defined as $\Lambda \equiv \sqrt{2\pi \beta \hbar^2/m}$ with $\beta$, $\hbar$ and $m$ denoting the inverse temperature, the reduced Planck constant, and the mass of the particles, respectively.
The collective variable $\bar{\mathbf r}$ represents the coordinates of the $3N$-dimensional configurational space: $\bar{\mathbf r} = \{{\mathbf r}_1, \cdots, {\mathbf r}_N
\}$. In Eq.~\eqref{eq:can}, each of the $3N$ variables spans a length of the period $L$.
The dependence of $Q_{\rm ex}$ on the volume $V=L^3$ arises from the prefactor $1/V^N$, the integration domain $L^{3N}$, and the potential energy ${\mathcal U}(\bar{\mathbf r}, L)$.
To simplify this dependence, scaled coordinates $\bar{\mathbf s}= \bar{\mathbf r}/L$ are introduced (e.g.\cite{Frenkel_Smit2023}), leading to the reformulation of $Q_{\rm ex}$ as
\begin{equation} Q_{\rm ex}(N,V,T) = \int_0^1 d\bar{\mathbf s}\, e^{-\beta {\mathcal U}(\bar{\mathbf s}L, L)} \label{eq:qex} .  \end{equation}
Here the scaled coordinates serve as dimensionless dummy variables, encapsulating the $V$-dependence entirely within the function ${\mathcal U}(\bar{\mathbf s}L, L)$.
Differentiating $Q(N,V,T)$ with respect to $V$ and subsequently differentiating Eq.~\eqref{eq:qex} define the thermodynamic pressure
\begin{equation} p \equiv \frac{1}{\beta} \left. \frac{\partial\log Q}{\partial V}\right|_{N,T} = \frac{N}{\beta V} + \frac{1}{3V} \left< A(\bar{\mathbf s},L) \right> ,  \label{eq:p} \end{equation}
where
\begin{equation} A(\bar{\mathbf s},L) = -L \cdot \left. \frac{\partial {\mathcal U}(\bar{\mathbf s}L, L)}{\partial L}\right|_{\bar{\mathbf s} } ,  \label{eq:A} \end{equation}
and the ensemble average $\left< A(\bar{\mathbf s},L)\right>$ is defined as a weighted integral over the scaled coordinates,
\begin{equation}  \left< A(\bar{\mathbf s},L) \right> \equiv  \frac{1}{Q_{\rm ex}} \int_0^1 d\bar{\mathbf s}\,  A(\bar{\mathbf s},L) e^{-\beta {\mathcal U}(\bar{\mathbf s}L, L)},  \label{eq:ena} \end{equation}
with the normalized weighting factor reflecting the Boltzmann distribution of configurations.
Clearly, the first term in the right hand side of Eq.~\eqref{eq:p} corresponds to the pressure of an ideal gas. To derive the second term, we use the relation $dL/dV = L/(3V)$.

Because ${\mathcal U}(\bar{\mathbf r}, L)$ depends on $L$ both explicitly and implicitly through $\bar{\mathbf r}=\bar{\mathbf s}L$, its derivative with respect to $L$ in Eq.~\eqref{eq:A} naturally splits into two contributions:
\begin{equation} A(\bar{\mathbf s},L) = \sum_{j=1}^N {\mathbf f}_j \cdot {\mathbf r}_j - L \left. \frac{\partial {\mathcal U}(\bar{\mathbf r}, L)}{\partial L}\right|_{\bar{\mathbf r} }  ,  \label{eq:pvirial}  \end{equation}
where ${\mathbf f}_j$ is the force acting on the $j$-th particle, and we have used the relation
\begin{equation} \left. \frac{ \partial{\mathbf r}_j}{\partial L} \right|_{ {\mathbf s}_j } = {\mathbf s}_j = \frac{ {\mathbf r}_j }{L} .  \label{eq:drl} \end{equation}
Notably, due to the explicit $L$-dependence of ${\mathcal U}(\bar{\mathbf r}, L)$, Eq.~\eqref{eq:pvirial} deviates from the standard virial expression by the correction term that typically accounts for the influence of periodic images\cite{Louwerse_Baerends2006,Thompson2009}.

The above Eqs.~\eqref{eq:can} to~\eqref{eq:drl} are valid for arbitrary periodic systems without making any assumption to the particular form of the energy.
For the system of the one-component plasma with the neutralizing background, if the underlying basic interaction is the Coulomb interaction or the angular-averaged interaction, both $\nu({\mathbf r},L)$ and $\tau([\nu])$ necessarily scales as $1/L$.
Consequently, the energy ${\mathcal U}(\bar{\mathbf r}, L)$, expressed as the combination of Eqs.~\eqref{eq:totalU},~\eqref{eq:Upp2}, and~\eqref{eq:ocp}, scales as $1/L$, leading to the derivative identity for ${\mathcal U}(\bar{\mathbf r}, L)$ as in Eq.~\eqref{eq:sb}.
$ A(\bar{\mathbf s},L)$ of Eq.~\eqref{eq:A} then reduces to ${\mathcal U}(\bar{\mathbf s}L, L)$ and thereby suggests a simple relation between the pressure and the ensemble average of the potential energy
\begin{equation} p = \frac{N}{\beta V} + \frac{1}{3V}  \left<  {\mathcal U}(\bar{\mathbf s}L, L)   \right> . \label{eq:pe} \end{equation}
According to the pairwise formulation of the energy, the justification of Eq.~\eqref{eq:pe} for the angular-averaged potential obviously arises from the fact that the cutoff distance $r_{\rm s}=(3/(4\pi))^{1/3}L$ depends linearly on $L$.
%\textcolor{blue}{
Any choice of a volume-dependent basic interaction in the form of Eq.~\eqref{eq:wr} where the parameter of the length scale is linearly proportional to $L$, will consistently validate Eq.~\eqref{eq:pe}.
In contrast, Eq.~\eqref{eq:pe} no longer holds for systems interacting via the custom basic interaction $w_{\rm cd}(r)$ whose length scale is independent of $L$.
This observation remains true even though the effective pairwise interaction under PBC is inherently volume-dependent.
%}

%%%%%%%%%%%%%%%%%%%%%%%%%%%%%%%%%%%%%%%%%%%%%%%%%%%%%%%%%%%%%%%%%%%%%%%%%%%%%%%%%%%%%%%%%%%%%%%%%%%%%%%%%%%%%%%%%%%%%%%%%%%%%%%%%%%%%%%%%%%%%%%%%%%%%%%%%%%%%%%%
\section{Conclusions}\label{sec:con}
We have developed a unified pairwise framework for transparently deriving the electrostatic energies of periodic systems that include both point charges and distributed charge densities.
This framework generalizes the earlier approach for systems of point charges \cite{Hu2014ib} and enables a straightforward and unambiguous extension of the Ewald formulation, as well as other related methods, to systems with arbitrary charge densities.

The application of the framework to the one-component plasma immediately demonstrates that the energy of the uniform background always vanishes, regardless of the specific treatment of the electrostatic interactions.
It not only clarifies the contributions arising from the background but also reveals the criteria necessary to preserve the simple energy-pressure relation for periodic Coulomb systems.

This general framework relies on the introduction of the effective pairwise interaction under the periodic boundary condition.
When the underlying basic interaction is the Coulomb interaction, a straightforward analysis of the infinite boundary term---defined as the conditional limit of  ${\mathbf k}\to{\mathbf 0}$\cite{Hu2014ib}---provides significant insights and yields a well-defined real-space series for the effective pairwise interaction,
$\nu_{\rm e3dtf}$. This effective pairwise interaction is now recognized as reflecting the bulk properties of an infinite crystal lattice.
The analysis applied to the Coulomb interaction can be readily extended to existing modified Coulomb interactions. The corresponding effective interactions exhibit universal properties that collectively characterize the effects of the periodic boundary condition.
Among these properties, bulk invariance and scaling behavior are unique to $\nu_{\rm e3dtf}$ and the effective interactions derived from certain modified Coulomb interactions.

The effective pairwise interaction $\nu({\mathbf r},L)$, is rigorously defined as a function of both the period $L$ and the relative vector ${\mathbf r}$.
%\textcolor{blue}{
Its clear physical interpretation, along with its explicit dependence on $L$, should enable a general treatment of finite-size effects, facilitating direct comparison with known results in the literature (e.g.\cite{Figueirido1995,Hummer1997,Remsing_Weeks2018}).
%}
As such, we ultimately hope that the unified formulation for energies can serve as a starting point, complementary to the traditional formulation, for addressing challenging problems related to other thermodynamic properties of periodic Coulomb systems
and for predicting structural and dielectric properties from the viewpoint of mean-field theories (e.g.\cite{Hu2014spmf,Hu2022,Remsing2023}).

\section*{Acknowledgement}
The authors have no conflicts to disclose. This work was supported by NSFC (Grant Nos. 22273047 and 21873037).

%%%%%%%%%%%%%%%%%%%%%%%%%%%%%%%%%%%%%%%%%%%%%%%%%%%%%%%%%%%%%%%%%%%%%%%%%%%%%%%%%%%%%%%%%%%%%%%%%%%%%%%%%%%%%%%%%%%%%%%%%%%%%%%%%%%%%%%%%%%%%%%%%%%%%%%%%%%%%%%%
\section*{Appendix: The Explicit Forms of the Pairwise Interactions}\label{sec:appA}
The effective pairwise interaction $\nu_{\rm e3dtf}$ [Eq.~\eqref{eq:e3dtf}] or equivalently $\nu_\infty$ [Eq.~\eqref{eq:ubulk}] can be decomposed into a sum of two absolutely and rapidly convergent series\cite{Hu2014ib,Yi_Hu2017pairwise},
\begin{equation} \nu_{\rm e3dtf}({\mathbf r},L) = \nu_\infty({\mathbf r},L) = \nu_{\rm R}({\mathbf r}) + \nu_{\rm F}({\mathbf r}), \label{eq:e3dtf2} \end{equation}
where the real-space series, which captures the near-field contribution of the Coulomb interaction, is given by
\begin{equation} \nu_{\rm R}({\mathbf r}) = \sum_{ {\mathbf n} }^{\infty} \frac{ {\rm erfc}(\alpha \lvert {\mathbf r}+ {\mathbf n}L \rvert) }{ \lvert {\mathbf r} + {\mathbf n}L \rvert }  - \sum_{ {\mathbf n}\neq 0 }^{\infty} \frac{ {\rm erfc}(\alpha  n L) }{ n L }  + \frac{2\alpha}{\sqrt{\pi}}, \label{eq:nuR} \end{equation}
and the Fourier-space series, which captures the far-field contribution of the Coulomb interaction, is expressed as
\begin{equation} \nu_{\rm F}({\mathbf r})  = \sum_{ {\mathbf n}\neq {\mathbf 0}}^\infty \frac{e^{-\pi^2n^2/(\alpha L)^2 }}{\pi n^2 L} \left( e^{i 2\pi {\mathbf n}\cdot {\mathbf r}/L} -1 \right)  .  \label{eq:nuF}  \end{equation}
Here, ${\rm erfc}(x)$ in Eq.~\eqref{eq:nuR} is the complementary error function. Eq.~\eqref{eq:e3dtf2} identifies with Eq.(3) of Ref.\cite{Yi_Hu2017pairwise}.
By appropriately choosing the parameter $\alpha > 0$, the computation of $\nu_{\rm e3dtf}({\mathbf r},L)$ via Eqs.~\eqref{eq:e3dtf2} to~\eqref{eq:nuF} becomes efficient for any ${\mathbf r}$.
Clearly, as $\alpha$ increases, the far-field contribution becomes more significant, while as $\alpha$ decreases, the near-field contribution dominates.

However, for convenience in deriving exact results, it is advantageous to express $\nu_{\rm e3dtf}$ entirely as a Fourier-space series. To achieve this, the real-space series in $\nu_{\rm R}({\mathbf r})$ is transformed into an equivalent Fourier-space representation,
%\begin{equation} 
\begin{multline} \sum_{ {\mathbf n} }^\infty  \frac{ {\rm erfc}(\alpha \lvert {\mathbf r}+ {\mathbf n}L \rvert) }{ \lvert {\mathbf r} + {\mathbf n}L \rvert }  =  \frac{\pi}{\alpha^2L^3} + \\
  \sum_{ {\mathbf n}\neq {\mathbf 0}}^\infty \frac{1-e^{-\pi^2n^2/(\alpha L)^2}}{\pi n^2 L} e^{i 2\pi {\mathbf n}\cdot {\mathbf r}/L } \label{eq:r}. \end{multline}
%\end{equation}
Consequently,  $\nu_{\rm R}({\mathbf r}) + \nu_{\rm F}({\mathbf r})$ reduces to Eq.~\eqref{eq:e3dtf}, i.e.,
\begin{equation} \nu_{\rm R}({\mathbf r}) + \nu_{\rm F}({\mathbf r}) = \frac{\xi}{L} + \sum_{ {\mathbf n}\neq {\mathbf 0}}^\infty \frac{e^{i 2\pi {\mathbf n}\cdot {\mathbf r}/L }}{\pi n^2 L} ,  \label{eq:e3dtf3}\end{equation}
where $\xi = 2.83\,729\,748\cdots$\cite{Placzek1951} is the constant independent of ${\mathbf r}$\cite{Yi_Hu2017pairwise},
%\begin{equation} \begin{aligned} \xi & = \frac{\pi}{\beta^2} + \frac{2\beta}{\sqrt{\pi}} - \sum_{ {\mathbf n}\neq 0 }^\infty \frac{ {\rm erfc}(\beta n) }{n} - \sum_{ {\mathbf n}\neq {\mathbf 0}}^{\infty} \frac{e^{-\pi^2n^2/\beta^2}}{\pi n^2}  \\
%     & = \lim_{s\to 0} \left[ \frac{1}{s} - \sum_{ {\mathbf n}\neq {\mathbf 0}}^{\infty} \frac{e^{i 2\pi {\mathbf n}\cdot {\mathbf s} }}{\pi n^2}  \right] .  \end{aligned} \end{equation}
\begin{equation} \xi = \frac{\pi}{\beta^2} + \frac{2\beta}{\sqrt{\pi}} - \sum_{ {\mathbf n}\neq 0 }^\infty \frac{ {\rm erfc}(\beta n) }{n} - \sum_{ {\mathbf n}\neq {\mathbf 0}}^{\infty} \frac{e^{-\pi^2n^2/\beta^2}}{\pi n^2}.  \end{equation}
Here we have introduced the dimensionless parameter $\beta = \alpha L > 0$, the dimensionless vector ${\mathbf s} = (s_1,s_2,s_3)$ and its magnitude $s = \lvert {\mathbf s} \rvert = \sqrt{s_1^2+s_2^2+s_3^2}$.
In deriving Eq.~\eqref{eq:e3dtf3}, no specific value is assumed for $\alpha$.
However, since the expression is now entirely written as a Fourier-space series, one may regard Eq.~\eqref{eq:e3dtf3} as being fully determined by the far-field contribution, corresponding to the limit $\alpha\to\infty$.

%\textcolor{blue}{
The ${\mathbf n}=0$ term, which is excluded from the summation in Eq.~\eqref{eq:nuF}, can be expanded to second order of ${\mathbf n}$  and identified with the infinite boundary term in Eq.~\eqref{eq:ib},
\begin{equation}  \begin{aligned} \nu_{\rm ib}({\mathbf r})  & = \frac{1}{2V} \lim_{{\mathbf k}\to 0}  \left[ - \frac{4\pi}{k^2} \left( {\mathbf k}\cdot{\mathbf r} \right)^2 \right] \\
  & = \frac{1}{2V} \lim_{\Omega\to\infty} \iiint_{\Omega} d{\mathbf x}\, \left( {\mathbf r} \cdot \nabla_{\mathbf x} \right)^2 \frac{1}{\left| {\mathbf x} \right|}       \end{aligned}, \label{eq:ib2} \end{equation}
where the change of variables ${\mathbf k}=2\pi {\mathbf n}/L$ and $V=L^3$ has been applied. Here, $\Omega$ is the volume of the crystal and $\nabla_{\mathbf x}$ denotes the gradient operator acting on the variable ${\mathbf x}$.
The second equality of Eq.~\eqref{eq:ib2} is obtained by interpretting the function inside the bracket as the three-dimensional fourier transform of the dipole-dipole interactions\cite{Pan2017}.
Using Gauss's divergence theorem, the above expression reduces to the familiar shape-dependent term (e.g.\cite{Smith2008})
\begin{equation}  \nu_{\rm ib}({\mathbf r}) = \frac{1}{2V} \lim_{\Omega\to\infty} \oiint_{\partial\Omega}  {\mathbf r}\cdot d{\mathbf S}  \left({\mathbf r}\cdot \nabla_{\mathbf x}\right) \frac{1}{\left| {\mathbf x} \right|}.  \end{equation}
This integral characterizes the summation order for an infinitely large crystal\cite{DeLeeuw_Smith1980,Smith2008}. However, its physical meaning may still remain unclear for a finite crystal, as discussed previously\cite{Hu2014ib}.
%}

The truncation schemes commonly employed in the literature\cite{Wolf1999,Fukuda2013,Wang_Fukuda2016} correspond to the following explicit expressions for $w_{\rm cd}(r)$,
\begin{equation} w_1(r) = \frac{{\rm erfc}(r/\sigma)}{r} , \label{eq:nvb1} \end{equation}
\begin{equation} w_2(r) = \frac{1}{r} - \frac{1}{8r_{\rm c}} \left(15- \frac{10 r^2}{r_{\rm c}^2}  + \frac{3 r^4}{r_{\rm c}^4}  \right) , \label{eq:nvb2} \end{equation}
and
\begin{equation} w_3(r) = \frac{1}{r} - \frac{1}{16r_{\rm c}} \left(35 - \frac{35 r^2}{r_{\rm c}^2} + \frac{21 r^4}{r_{\rm c}^4} -\frac{5 r^6}{r_{\rm c}^6}  \right) , \label{eq:nvb3}  \end{equation}
where $\sigma$ is the screening length and $r_{\rm c}$ is the cutoff distance. Both $\sigma$ and $r_{\rm c}$ play the role of the length scales for $w_j(r)$.
For a fixed dimensionless parameter $s$, it is evident that $w_1(s\sigma)$, $w_2(sr_{\rm c})$, and $w_3(sr_{\rm c})$ all scale inversely with their length scales.
The three-dimensional Fourier transforms of these functions are related to the Fourier transform of the Coulomb interaction via
\begin{equation} \hat{w}_j({\mathbf k}) = \frac{4\pi}{k^2} \left[ 1 + \hat{d}_j({\mathbf k}) \right], \end{equation} where
\begin{equation} \hat{d}_1({\mathbf k}) = - e^{-k^2\sigma^2/4},  \end{equation}
\begin{equation} \hat{d}_2({\mathbf k}) = 15 \frac{ 3k_{\rm c}\cos k_{\rm c} - (3-k_{\rm c}^2)\sin k_{\rm c} }{k_{\rm c}^5} ,  \label{eq:d2} \end{equation}
and
\begin{equation} \hat{d}_3({\mathbf k}) = 105\frac{(15-k_{\rm c}^2)k_{\rm c} \cos k_{\rm c} - (15-6k_{\rm c}^2)\sin k_{\rm c} }{k_{\rm c}^7}. \label{eq:d3} \end{equation}
Here, $k_{\rm c} = k r_{\rm c}$ in Eqs.~\eqref{eq:d2} and~\eqref{eq:d3}.
These Fourier transforms have been derived previously\cite{Hu2022}. See Eqs.(7) to (10) of Ref.\cite{Hu2022}. Note that a prefactor $1/(4\pi\varepsilon_0)$ should be included to account for the difference between Gaussian units and SI units. 
Under PBC, the effective pairwise interactions follow Eq.~\eqref{eq:wcdpbc},
\begin{equation} \nu_j({\mathbf r},L) = \tau_j   +  \frac{1}{L^3}\sum_{ {\mathbf n}\neq{\mathbf 0} }^\infty  e^{i 2\pi {\mathbf n}\cdot {\mathbf r}/L }\hat{w}_j(2\pi{\mathbf n}/L) ,  \label{eq:wcdpbc2} \end{equation}
where the constant term $\tau_j$ is given by
\begin{equation} \tau_j =\lim_{r\to 0 } \left[\frac{1}{r} - w_j(r) \right]   + \frac{1}{L^3} \lim_{{\mathbf k}\to{\mathbf 0}} \hat{w}_j({\mathbf k}), \end{equation}
yielding explicitly,
\begin{equation} \tau_1 = \frac{2}{\sqrt{\pi}\sigma} + \frac{\pi\sigma^2}{L^3},    \end{equation}
and
\begin{equation} \tau_2 = \frac{2\pi r_{\rm c}^2}{7L^3}+\frac{15}{8r_{\rm c}}; \quad \tau_3 = \frac{2\pi r_{\rm c}^2}{9L^3}+\frac{35}{16r_{\rm c}}. \end{equation}
These constant terms play a crucial role in ensuring that the effective pairwise interactions dominate over the bare Coulomb interaction. The behaviors of $\nu_j({\mathbf r},L)$ are qualitatively similar to those shown in Fig.~\ref{fig:e3dtf}.
%\textcolor{blue}{
Notably, none of the effective interactions---$\nu_1({\mathbf r},L)$, $\nu_2({\mathbf r},L)$, and $\nu_3({\mathbf r},L)$---exhibits scaling behavior [Eq.~\eqref{eq:sb}] under the condition that the parameters $\sigma$ and $r_{\rm c}$ are fixed and independent of $L$.
However, if these length scales were assumed to be linearly proportional to $L$, the scaling property of Eq.~\eqref{eq:sb} would be restored.
%}

%\textcolor{blue}{
The effective pairwise interaction expressed as the fourier series [Eqs.~\eqref{eq:e3dtf},~\eqref{eq:wpbc},~\eqref{eq:wcdpbc} and ~\eqref{eq:wcdpbc2}] always takes the form
\begin{equation} \nu({\mathbf r},L) = \tau + \sum_{{\mathbf n}\neq 0} f({\mathbf n}) \cos \frac{2\pi {\mathbf n}\cdot {\mathbf r}}{L} , \end{equation}
where $f({\mathbf n})$ is an even function of ${\mathbf n}=(n_1,n_2,n_3)$. Under PBC, the electric field at the surface of the primary cell generated by a source point charge located at the center of the cell is given by
%\begin{equation}
\begin{multline}  -{\mathbf e}_x \cdot \left. \nabla \nu({\mathbf r}, L)\right|_{x=\pm L/2}  = \\ \sum_{{\mathbf n}\neq 0} f({\mathbf n})\frac{2\pi}{L} n_1 \cos(n_1 \pi) \sin\frac{2\pi (n_2y + n_3 z) }{L}  \end{multline}
%\end{equation}
in the $x$-direction. This field necessarily vanishes because the summation involves an odd function of $n_1$, ensuring cancellation over symmetric terms. Thus, property (iv) of Eq.~\eqref{eq:ve} is proven.
%}
%%%%%%%%%%%%%%%%%%%%%%%%%%%%%%%%%%%%%%%%%%%%%%%%%%%%%%%%%%%%%%%%%%%%%%%%%%%%%%%%%%%%%%%%%%%%%%%%%%%%%%%%%%%%%%%%%%%%%%%%%%%%%%%%%%%%%%%%%%%%%%%%%%%%%%%%%%%%%%%%

%\bibliography{refsall}

\end{document}